\documentclass[sigconf]{acmart} 
\AtBeginDocument{%
  }

\usepackage{multirow}
\usepackage{enumitem}

\copyrightyear{2025}
\acmYear{2025}
\setcopyright{cc}
\setcctype{by}
\acmConference[CIKM '25]{Proceedings of the 34th ACM International Conference on Information and Knowledge Management}{November 10--14, 2025}{Seoul, Republic of Korea}
\acmBooktitle{Proceedings of the 34th ACM International Conference on Information and Knowledge Management (CIKM '25), November 10--14, 2025, Seoul, Republic of Korea}
\acmDOI{10.1145/3746252.3760911}
\acmISBN{979-8-4007-2040-6/2025/11}

\begin{document}

\title{A Universal Framework for Offline Serendipity Evaluation in Recommender Systems via Large Language Models}

\author{Yu Tokutake}
\authornote{Corresponding author.}
\orcid{0009-0007-1399-1631}
\affiliation{%
  \institution{The University of Electro-Communications}
  \city{Chofu}
  \state{Tokyo}
  \country{Japan}
}
\email{tokutakeyuu@uec.ac.jp}

\author{Kazushi Okamoto}
\authornotemark[1]
\orcid{0000-0002-9571-8909}
\affiliation{%
  \institution{The University of Electro-Communications}
  \city{Chofu}
  \state{Tokyo}
  \country{Japan}}
\email{kazushi@uec.ac.jp}

\author{Kei Harada}
\orcid{0009-0009-5453-6699}
\affiliation{%
  \institution{The University of Electro-Communications}
  \city{Chofu}
  \state{Tokyo}
  \country{Japan}
}
\email{harada@uec.ac.jp}

\author{Atsushi Shibata}
\orcid{0000-0003-1794-8562}
\affiliation{%
  \institution{Advanced Institute of Industrial Technology}
  \city{Shinagawa-ku}
  \state{Tokyo}
  \country{Japan}
}
\email{shibata-atsushi@aiit.ac.jp}

\author{Koki Karube}
\orcid{0009-0005-2699-1193}
\affiliation{%
  \institution{The University of Electro-Communications}
  \city{Chofu}
  \state{Tokyo}
  \country{Japan}
}
\email{karubekoki@uec.ac.jp}


\begin{abstract}
  Serendipity in recommender systems (RSs) has attracted increasing attention as a concept that enhances user satisfaction by presenting unexpected and useful items.
However, evaluating serendipitous performance remains challenging because its ground truth is generally unobservable.
The existing offline metrics often depend on ambiguous definitions or are tailored to specific datasets and RSs, thereby limiting their generalizability.
To address this issue, we propose a universally applicable evaluation framework that leverages large language models (LLMs) known for their extensive knowledge and reasoning capabilities, as evaluators.
First, to improve the evaluation performance of the proposed framework, we assessed the serendipity prediction accuracy of LLMs using four different prompt strategies on a dataset containing user-annotated serendipitous ground truth and found that the chain-of-thought prompt achieved the highest accuracy.
Next, we re-evaluated the serendipitous performance of both serendipity-oriented and general RSs using the proposed framework on three commonly used real-world datasets, without the ground truth.
The results indicated that there was no serendipity-oriented RS that consistently outperformed across all datasets, and even a general RS sometimes achieved higher performance than the serendipity-oriented RS.

\end{abstract}

\ccsdesc[500]{Information systems~Recommender systems}

\keywords{Serendipity, Recommender Systems, Large Language Models, Evaluation}


\maketitle

\section{Introduction} \label{sec:introduction}
Recommender systems (RSs) play a crucial role in supporting users' decision-making by providing personalized recommendations in the era of information overload~\cite{2023aug_z.Fu}.
Traditional RSs, such as collaborative filtering, primarily focus on recommendation accuracy by modeling user-item interactions from historical behaviors.
However, such approaches often cause over-specialization, leading to the filter bubble~\cite{2023aug_z.Fu, 2023apr_z.Li, 2025feb_y.Xi}, where users encounter limited items similar to past preferences.
To address this issue, the concept of serendipity in RSs was proposed~\cite{2021mar_r.j.Ziarani, 2023aug_z.Fu, 2004jan_j.Herlocker}.
Although there is no consensus on a clear definition, many studies~\cite{2023jul_z.Fu, 2021oct_m.Zhang, 2021dec_r.j.Ziarani, 2024aug_z.Fu} have considered serendipity in RSs to comprise two main components: relevance and unexpectedness.
Evaluating serendipity is also challenging because of its subjective nature and the lack of ground truth.
Ideally, serendipity should be evaluated through user studies~\cite{2023jul_z.Fu, 2018apr_d.Kotkov, 2019may_l.Chen, 2020jul_n.Wang}; however, owing to practical constraints, researchers often rely on offline evaluations based on historical user behavior.

Existing offline serendipity evaluation methods can be categorized into (1) methods that define custom metrics and (2) methods that use datasets with serendipitous ground truths.
For (1), various metrics have been proposed to identify items that are both relevant and unexpected to a user and then measure the overlap between those items and the recommendation list~\cite{2007jun_t.Murakami, 2017sep_d.Kotkov, 2015jun_q.Zheng}.
Although intuitive and straightforward, these metrics typically depend on predefined thresholds for relevance and unexpectedness.
For example, Zheng et al.~\cite{2015jun_q.Zheng} classified unexpected items as those that were not in the top 50 most highly rated or the top 50 most popular items in the test set.
However, this definition assumes the same threshold for all users and may not be appropriate for every scenario.
Moreover, some studies~\cite{2020sep_p.Li, 2021oct_m.Zhang, 2023oct_y.Tokutake} have evaluated serendipity using RS-specific metrics, hindering fair comparisons across different RSs.
In (2), datasets with serendipitous ground truths enable evaluation using standard accuracy-based metrics (e.g., precision and normalized discounted cumulative gain (NDCG))~\cite{2021dec_r.j.Ziarani, 2023jul_z.Fu, 2024aug_z.Fu, 2025feb_y.Xi}.
However, datasets such as Serendipity-2018~\cite{2018apr_d.Kotkov} and SerenLens~\cite{2023jul_z.Fu} focus only on specific domains (movies and books), thereby restricting their broader applicability.

\begin{figure}[t]
  \centering
  \includegraphics[width=0.9\linewidth]{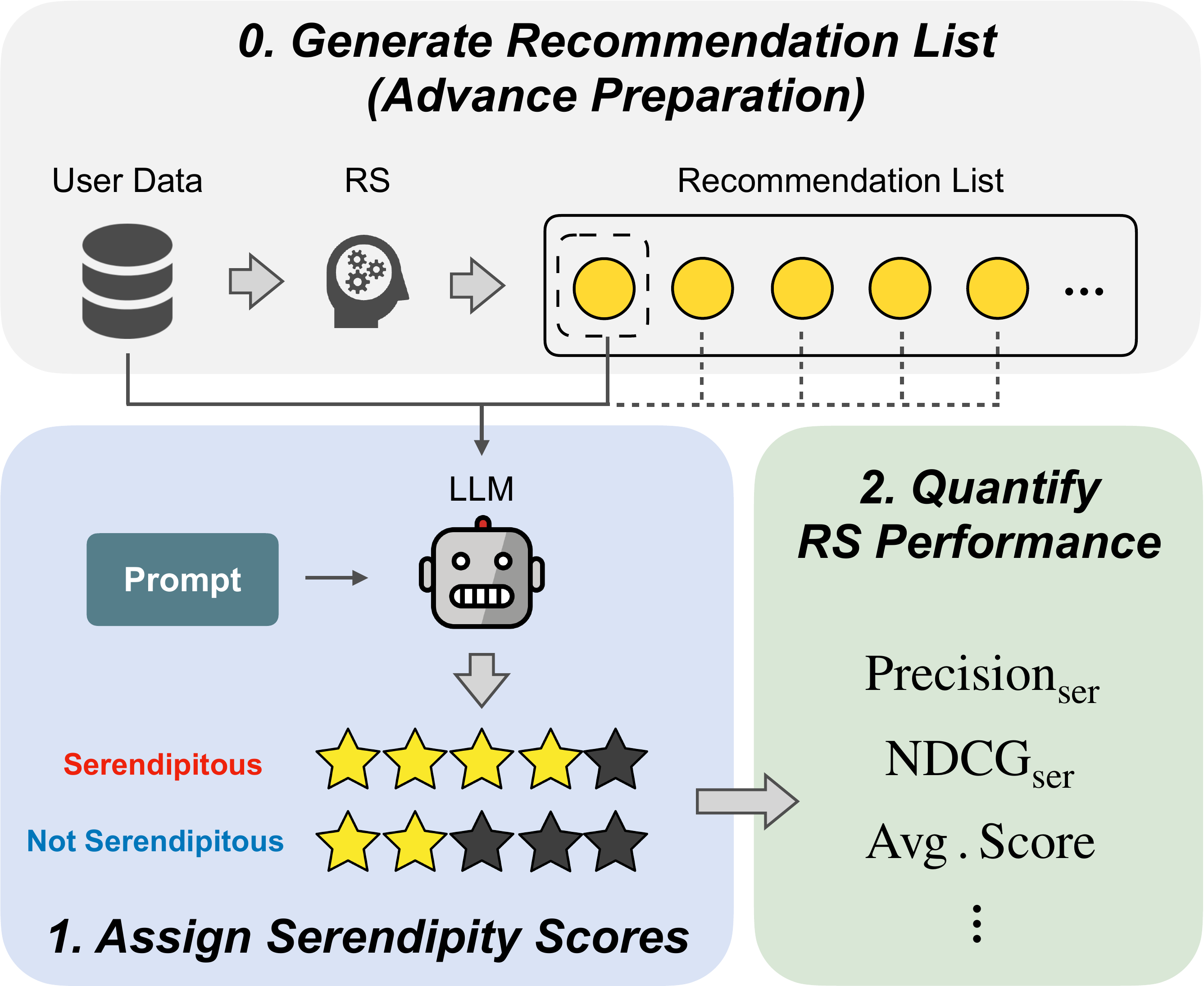}
  \caption{Overview of the proposed evaluation framework. Items with a serendipity score above four are considered serendipitous to the user.}
  \vspace{-4mm}
  \label{fig:proposed_framework}
\end{figure}

Recently, the concept of ``LLM-as-a-Judge'' has drawn increasing attention~\cite{2024dec_j.Gu, 2023dec_l.Zheng, 2023aug_g.Faggioli}.
This approach leverages large language models (LLMs), known for their human-like reasoning capabilities, as evaluators that combine human-like contextual reasoning with the scalability of automatic metrics.
Because LLMs are pre-trained on large and diverse text corpora, they can be evaluated using extensive domain knowledge.
Tokutake et al.~\cite{2024nov_y.Tokutake} investigated whether LLMs could judge whether an item is serendipitous to a user by comparing their judgments with the ground truth labels in the Serendipity-2018 dataset.
They found the best performance when item names and categories were provided as input to LLMs and that the LLM-as-a-Judge approach outperformed the baseline methods, although there remains room for improvement.
However, to the best of our knowledge, no study has utilized LLM-as-a-Judge to evaluate serendipitous performance in RSs.

In this study, we propose a universal offline serendipity evaluation framework that leverages the LLM-as-a-Judge.
The only input to the framework is textual information about the user and the recommended item.
Therefore, our approach requires neither ambiguous definitions nor a specific RS and is not limited to datasets with serendipitous ground truth.
Specifically, the framework consists of two steps:
(1) assigning serendipity scores to items recommended to users using the LLM-as-a-Judge system; and
(2) quantifying serendipitous performance based on these scores.
To manage diverse evaluation scenarios, our LLM-as-a-Judge system provides a five-level serendipity score for each recommended item, offering greater granularity than the binary approach used in a previous study~\cite{2024nov_y.Tokutake}.
Once these scores are assigned, RSs can be evaluated either directly using the score values themselves or by creating a serendipitous ground truth from these scores.
Examples of the former include metrics such as the average score within the recommendation list, whereas the latter can apply accuracy-based metrics.
In addition, because the proposed framework assigns scores to all recommended items, it supports counterfactual evaluations, even for items with which users have not actually interacted.
Figure \ref{fig:proposed_framework} illustrates our proposed framework.

To verify the effectiveness of the proposed framework, we conducted two experiments.
First, we performed a prompt selection experiment to identify the most effective prompting strategy for the LLM-as-a-Judge system.
We prepared four candidate prompt templates and assessed their accuracy in predicting serendipity using the Serendipity-2018 dataset.
Second, we conducted a comprehensive serendipity evaluation using the proposed framework with the previously identified best-performing prompt strategy.
Specifically, we re-evaluate two accuracy-oriented RSs and four serendipity-oriented RSs on three datasets without serendipitous ground truths.
In summary, our main contributions are as follows:
\begin{itemize}
  \item We present a novel offline serendipity evaluation framework leveraging LLM-as-a-Judge.
  \item We demonstrated that our framework is universally applicable and independent of specific datasets and RSs.
  \item We evaluated both the serendipity judgment capability of LLMs and the serendipitous performance of RSs using the proposed framework.
\end{itemize}
\section{Related Work} \label{sec:related_work}
\textbf{Serendipity-oriented RSs. }
Serendipity-oriented RSs aim to enhance serendipity by incorporating serendipity-aware components into various algorithms, including $k$-nearest neighbor ($k$-NN)~\cite{2013feb_a.Said}, matrix factorization (MF)~\cite{2015jun_q.Zheng}, and deep learning (DL)~\cite{2020apr_x.Li, 2020sep_p.Li, 2021dec_r.j.Ziarani, 2023jul_z.Fu, 2021oct_m.Zhang}.
Recently, serendipity-oriented RSs that utilize LLMs have been proposed~\cite{2024aug_z.Fu, 2025feb_y.Xi}.
Although similar to our approach in that these studies use LLMs to estimate serendipity scores for user-item pairs, our proposed framework differs by employing LLMs as evaluators rather than as recommendation models themselves.

\noindent
\textbf{LLM-as-a-Judge. }
LLM-as-a-Judge has been adopted as an evaluation mechanism for a wide range of tasks.
Many studies have demonstrated that evaluations using LLM-as-a-Judge are highly correlated with human judgments~\cite{2023aug_g.Faggioli, 2023dec_l.Zheng}, reinforcing its reliability.
In the field of RSs, LLM-as-a-Judge has been applied to evaluate recommendation explanations~\cite{2024oct_x.Zhang} and to compare recommendation lists~\cite{2024dec_z.Wu}.
However, a comprehensive serendipity evaluation framework for RSs using LLM-as-a-Judge is yet to be proposed.
In addition, the most direct and effective way to enhance LLM-as-a-Judge evaluation performance is to optimize the prompt design~\cite{2024dec_j.Gu}.
Therefore, the first step in this study was to develop a prompt strategy to improve the accuracy of serendipity predictions.

\section{LLM-as-a-Judge System Construction} \label{sec:proposed_framework}

\subsection{Problem Formulation}
The proposed LLM-as-a-Judge system takes as input the user's last $n$ historical behaviors $I_{u, n}$ and a recommended item $i$, and provides a five-level serendipity score $\mathcal{E}_{ui}$.
Formally, the score is given by
\begin{equation}
  \mathcal{E}_{ui} \leftarrow \mathcal{P}_\mathcal{LLM} (x(I_{u, n}, i) \oplus \mathcal{C}), \quad  \mathcal{E}_{ui} \in \{1, 2, 3, 4, 5\}
\end{equation}
where $\mathcal{P}_\mathcal{LLM}$ denotes a probability function defined by the LLM, $\oplus$ is the concatenation, and $\mathcal{C}$ is a prompt that provides additional context to the LLM.
A score of $\mathcal{E}_{ui} = 3$ indicates neutrally, with lower values (toward 1) denoting less serendipitous and higher values (toward 5) denoting more serendipitous items.
Following a previous study~\cite{2024nov_y.Tokutake}, we set $n = 10$ and defined $x(I_{u, n}, i)$ as the concatenation of item names and categories from $I_{u, n}$ and $i$.

\subsection{Prompt Design}
\begin{figure}[t]
  \centering
  \includegraphics[width=\linewidth]{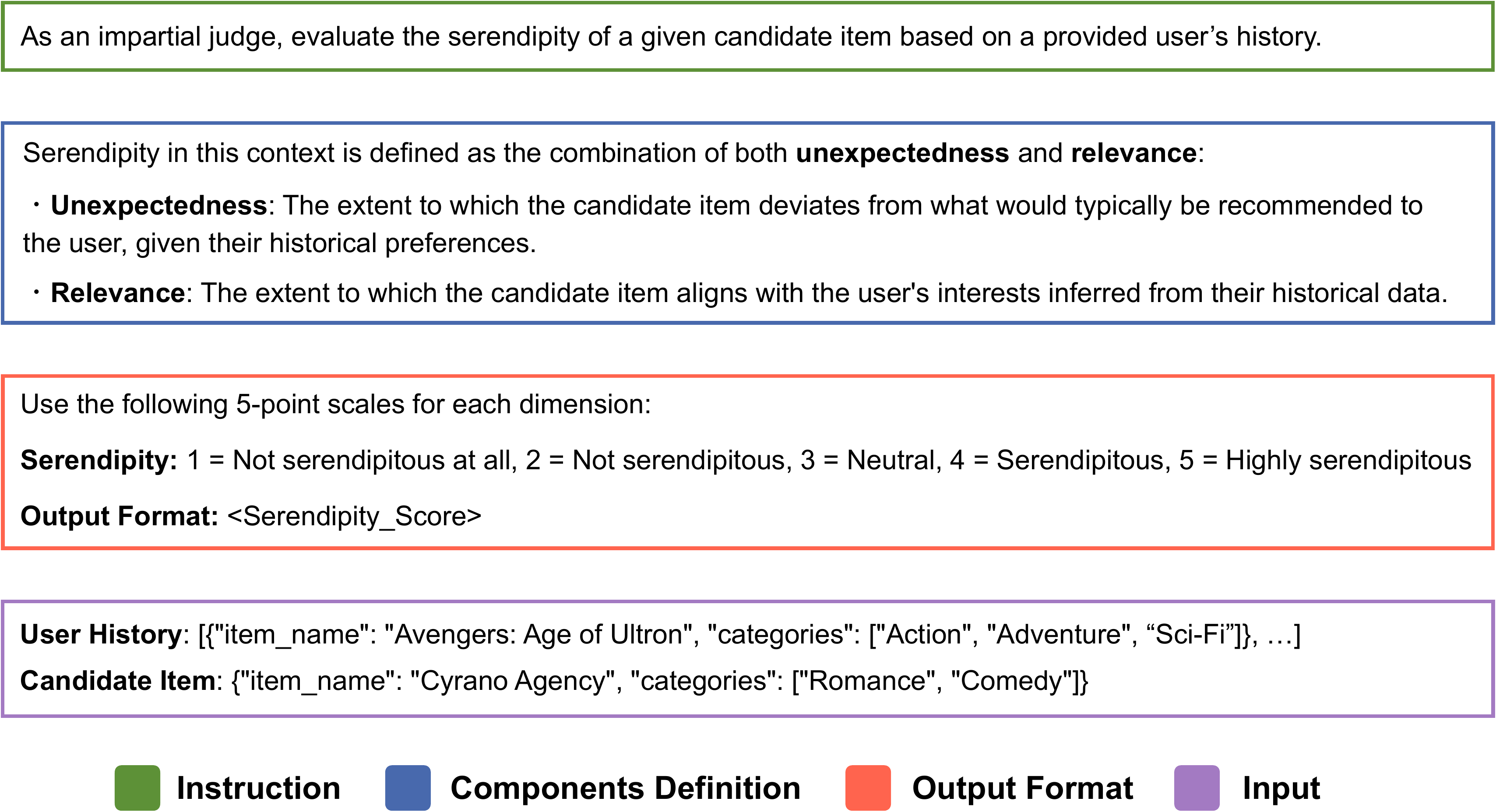}
  \caption{Example of a prompt template}
  \Description{aa}
  \label{fig:prompt_template}
  \vspace{-3mm}
\end{figure}

As in many previous studies~\cite{2023jul_z.Fu, 2021oct_m.Zhang, 2021dec_r.j.Ziarani, 2024aug_z.Fu}, we define serendipity as a combination of relevance and unexpectedness.
Based on this definition, we propose four prompt templates that employ common prompt-engineering strategies.
Figure~\ref{fig:prompt_template} illustrates the overall template, which combines the core components and inputs common to all prompts.
Descriptions of each prompt type are provided below:
\begin{itemize}
  \item \textbf{Base}: A prompt that outputs a five-level serendipity score. The scale shown in Figure~\ref{fig:prompt_template} is not included in this prompt.
  \item \textbf{LS}: Prompt extending the base prompt by including a Likert scale description of the five-level output.
  \item \textbf{CoT}: A prompt that incorporates chain-of-thought reasoning~\cite{2023jan_j.Wei} into the LS prompt. Specifically, it breaks down the evaluation into three steps: relevance, unexpectedness, and serendipity, outputting individual scores for each step.
  \item \textbf{LtM}: Prompt further subdivision of the three-step CoT approach into separate subprompts following the least-to-most strategy~\cite{2023may_d.Zhou}. The output of each sub-prompt serves as the input for the next, requiring three LLM responses per item.
\end{itemize}
For all the prompt templates, we employed a 10-shot prompting approach, providing two concrete examples for each serendipity score levels ($\mathcal{E} = 1, 2, 3, 4, 5$).

\section{Prompt Selection Experiment} \label{sec:prompt_selection_exp}
Using a dataset containing serendipitous ground truth, we evaluated our LLM-as-a-Judge system using four prompt templates.

\subsection{Experimental Setup}
\textbf{Dataset. }
The Serendipity-2018 dataset~\cite{2018apr_d.Kotkov} was used in this experiment.
The five-level serendipitous ground truth $g$ is defined as the average of the five-level responses to the six serendipity questions in the dataset rounded to the nearest integer.
This process yielded the following distributions of $g$: 1 (321), 2 (1,022), 3 (586), 4 (208), and 5 (13).
The SerenLens dataset~\cite{2023jul_z.Fu}, on which a user study was conducted, was not applicable in this experiment because it provides only a binary serendipitous ground truth.

\noindent
\textbf{LLMs. }
We employed two LLMs for the LLM-as-a-Judge system: GPT-4o-mini~\cite{2024oct_openai} (GPT) and the 4-bit quantization version of Llama-3.1-70B Instruct~\cite{2024nov_a.grattafiori} (Llama).
For both models, the temperature was set to 0.0 to ensure that the responses were as consistent as possible.

\noindent
\textbf{Evaluation Metrics. }
Our goal was to assess the accuracy of the LLM-as-a-Judge system in predicting the five-level serendipity scores using each prompt template.
Therefore, we calculated the mean absolute error (MAE) between the predicted scores $\mathcal{E}$ and ground truth labels $g$.
Additionally, we computed the three-class accuracy by categorizing both the predicted and ground truth scores into three groups: negative ($<3$), neutral ($=3$), and positive ($>3$).

\begin{table}[t]
  \caption{
    Prediction performance for each prompt template.
    The best and second-best results are bold and underlined, respectively.
    Accuracy is reported as a percentage.
  }
  \label{tab:serendipity_prediction_performance}
  \begin{tabular}{ccccc}
    \toprule
    \multirow{2}{*}{Prompt Template} & \multicolumn{2}{c}{MAE} & \multicolumn{2}{c}{Accuracy} \\
    \cmidrule(lr){2-3} \cmidrule(lr){4-5}
    & GPT & Llama & GPT & Llama \\
    \midrule
    Base & 1.473 & 2.265  & 20.79 & 27.49 \\
    LS & 1.159 & 1.731 & 30.74 & 13.67 \\
    CoT & \textbf{0.8470} & \textbf{1.117} & \textbf{46.98} & \underline{34.05} \\
    LtM & \underline{1.014} & \underline{1.356} & \underline{40.88} & \textbf{35.26} \\
    \bottomrule
  \end{tabular}
\end{table}

\begin{table}[t]
  \caption{Statistics of datasets used for RSs evaluation}
  \label{tab:dataset_statistics}
  \begin{tabular}{ccccc}
    \toprule
    Dataset & Users & Items & Interactions & Categories \\
    \midrule
    ML-1M & 6,038 & 3,533 & 575,281 & 20 \\
    Goodreads & 18,522 & 25,439 & 883,573 & 11 \\
    Beauty & 22,269 & 12,086 & 154,272 & 233 \\
    \bottomrule
  \end{tabular}
\end{table}

\begin{table*}[t]
  \centering
  \caption{%
    Performance comparison of RSs on three datasets.
    The best results are shown in bold, and the second-best results are underlined.
    $\text{P}_\mathrm{acc}$, $\text{N}_\mathrm{acc}$, $\text{P}_\mathrm{ser}$, $\text{N}_\mathrm{ser}$, and Avg denote $\text{Precision}_\mathrm{acc}$@20, $\text{NDCG}_\mathrm{acc}$@20, $\text{Precision}_\mathrm{ser}$@20, $\text{NDCG}_\mathrm{ser}$@20, and Avg.\ Score@20, respectively.
    All results are presented as three-fold averages; values are expressed as percentages, except for Avg.
  }
  \label{tab:rs_performance}
  \begin{tabular}{cccccccccccccccc}
    \toprule
    \multirow{2}{*}{} & \multicolumn{5}{c}{ML-1M} & \multicolumn{5}{c}{Goodreads} & \multicolumn{5}{c}{Beauty} \\
    \cmidrule(lr){2-6} \cmidrule(lr){7-11} \cmidrule(lr){12-16}
    & P$_\mathrm{acc}$ & N$_\mathrm{acc}$ & P$_\mathrm{ser}$ & N$_\mathrm{ser}$ & Avg & P$_\mathrm{acc}$ & N$_\mathrm{acc}$ & P$_\mathrm{ser}$ & N$_\mathrm{ser}$ & Avg & P$_\mathrm{acc}$ & N$_\mathrm{acc}$ & P$_\mathrm{ser}$ & N$_\mathrm{ser}$ & Avg \\
    \midrule
    BPRMF  & \underline{8.20} & \underline{9.34} & \textbf{8.51} & \underline{8.17} & \underline{2.378} & 0.828 & 1.41 & 0.995 & 0.991 & 2.131 & \textbf{0.515} & \textbf{1.77} & \underline{0.412} & 0.389 & \underline{2.189} \\
    SASRec & 6.32 & 8.11 & 5.95 & 5.87 & 2.339 & \textbf{1.26} & \textbf{2.76} & 1.28 & 1.30 & 2.199 & 0.135 & 0.642 & 0.243 & 0.210 & 2.187 \\
    \midrule
    KFN  & 1.32 & 1.39 & 3.38 & 3.36 & 2.134 & 0.0417 & 0.101 & 1.13 & 1.19 & \textbf{2.294} & 0.0267 & 0.0846 & 0.125 & 0.105 & 2.168 \\
    UAUM & 7.09 & 7.84 & \underline{8.25} & 8.16 & 2.334 & \underline{0.913} & 1.54 & \textbf{1.41} & \underline{1.38} & 2.156 & 0.173 & 0.685 & 0.393 & \underline{0.398} & 2.155 \\
    DESR & 4.00 & 4.39 & 7.90 & \textbf{8.38} & 2.346 & 0.0717 & 0.114 & 1.01 & 1.05 & \underline{2.270} & 0.0183 & 0.0980 & 0.152 & 0.145 & 2.172 \\
    PURS & \textbf{8.28} & \textbf{11.8} & 6.50 & 6.59 & \textbf{2.387} & 0.908 & \underline{1.81} & \underline{1.33} & \textbf{1.41} & 2.228 & \underline{0.283} & \underline{1.03} & \textbf{1.06} & \textbf{0.946} & \textbf{2.224} \\
    \bottomrule
  \end{tabular}
\end{table*}

\subsection{Results and Discussion}
Table \ref{tab:serendipity_prediction_performance} presents the prediction performance of the LLM-as-a-Judge system using each proposed prompt templates for the two LLMs.
Except for the accuracy using Llama, CoT achieved the best results across the LLM and the metrics.
In particular, when using the GPT, the CoT achieved an MAE below 1.0, indicating that the system can effectively distinguish between serendipitous ($\mathcal{E} > 3$) and non-serendipitous ($\mathcal{E} < 3$) items on average.
LtM attained the second-best performance, suggesting that stepwise inference prompts (CoT and LtM) significantly improved the prediction accuracy of the LLM-as-a-Judge system.

\section{RSs Performance Evaluation Experiment} \label{sec:rs_evaluation_exp}
We re-evaluated several RSs using our proposed framework.
Based on the results presented in Section~\ref{sec:prompt_selection_exp}, we adopted the GPT with the CoT prompt as the LLM-as-a-Judge system because it provided the best prediction performance.
All other settings for the LLM-as-a-Judge were consistent with those used in previous experiments.

\subsection{Experimental Setup}
\textbf{Dataset. }
We used three real-world datasets that are commonly employed in RS studies: MovieLens-1M~\cite{2015dec_f.Harper} (ML-1M) from the movie domain, Goodreads~\cite{2019jul_m.Wan} from the book domain, and the Beauty subset of Amazon reviews~\cite{2016apr_r.He} from the e-commerce domain.
Each dataset contains five-level user ratings but no serendipitous ground truth.
To convert the data into implicit feedback, all interactions with ratings of four or higher were treated as positive, and all other pairs were discarded.
Negative feedback was sampled randomly.
Table \ref{tab:dataset_statistics} lists the statistics of these datasets.

Three-fold cross-validation~\cite{2023feb_m.Andronov} was performed to evaluate the generalization performance.
To prevent data leakage, we used a temporal global split~\cite{2020sep_z.Meng}: we sorted the interactions chronologically, and partitioned them into 10 equally sized subsets.
The first six subsets served as training, the next subset as validation, and the subsequent subset as testing.
We repeated this process three times, each time the starting subset was shifted by one.

\noindent
\textbf{Evaluation Metrics. }
We evaluated RS performance in terms of recommendation accuracy and as well as serendipity using the proposed framework.
We used $\text{Precision}_\textrm{acc}$@20~\cite{2021oct_m.Zhang} and $\text{NDCG}_\textrm{acc}@20$~\cite{2002oct_k.Jarvelin} based on the positive interactions in the test set for accuracy metrics.
For serendipitous metrics, we employed $\text{Precision}_\textrm{ser}@20$~\cite{2021dec_r.j.Ziarani} and $\text{NDCG}_\textrm{ser}@20$~\cite{2023jul_z.Fu}, where items receiving an LLM-as-a-Judge score $\mathcal{E}$ of four or higher were considered serendipitous.
Additionally, we calculated the Avg.\ Score@20 represents the average $\mathcal{E}$ of the top-20 recommended items.

\noindent
\textbf{Comparison of RSs. }
We compared six RSs: two well-known accuracy-oriented and four serendipity-oriented RSs.
For the accuracy-oriented RSs, \textbf{BPRMF}~\cite{2009jun_s.Rendle} and \textbf{SASRec}~\cite{2018jun_s.Kang} were selected.
The four serendipity-oriented RSs included:
\begin{itemize}
  \item \textbf{KFN}~\cite{2013feb_a.Said}: A $k$-NN-based method recommending items disliked by users with dissimilar preferences.
  \item \textbf{UAUM}~\cite{2015jun_q.Zheng}: A MF-based approach incorporating an unexpectedness term into singular value decomposition.
  \item \textbf{DESR}~\cite{2020apr_x.Li}: A DL model infering users' long-term preferences and short-term demands to capture serendipity.
  \item \textbf{PURS}~\cite{2020sep_p.Li}: A DL model capturing unexpectedness in user sessions using a self-attention mechanism.
\end{itemize}

We optimized the hyperparameters of each RS using the Tree-structured Parzen Estimator~\cite{2023may_s.Watanabe} with 100 trials per RS per fold.
$\text{Recall}_\textrm{acc}$@10~\cite{2023oct_y.Tokutake} was used as the objective function to ensure sufficient recommendation accuracy.
Although state-of-the-art serendipity-oriented RSs may achieve a higher serendipitous performance, they require multiple sessions per user~\cite{2021oct_m.Zhang} or serendipitous ground truth for training~\cite{2023jul_z.Fu, 2024aug_z.Fu, 2025feb_y.Xi}, making them unsuitable for our setting.

\subsection{Results and Analysis}
Table~\ref{tab:rs_performance} shows the performance of each RS.
PURS ranked among the top on most serendipitous metrics across datasets (except for $\text{P}_\textrm{ser}$ and $\text{N}_\textrm{ser}$ in ML-1M and Avg in Goodreads), the other RSs performed well in specific scenarios.
Our key observations are as follows:

\noindent
\textbf{(1) Relationship between the accuracy and serendipitous performance. }
No RS achieved the best performance across all metrics.
However, PURS and UAUM on Goodreads, and BPRMF on ML-1M and Beauty, showed strong performances in both accuracy and serendipity.
These findings suggest that serendipity detection can be enhanced by maintaining a reasonable level of accuracy.

\noindent
\textbf{(2) Differences in serendipitous performance across RSs and datasets. }
Although the PURS generally indicated a high serendipitous performance, the BPRMF also performed well in ML-1M and Beauty.
Meanwhile, UAUM ranked first or second on at least one serendipitous metric for each dataset.
These outcomes indicate that serendipity-oriented RSs do not always outperform accuracy-oriented ones and that a simple RS such as UAUM can sometimes surpass more complex DL models.
Furthermore, the performance varies across datasets (domains), suggesting that the most suitable serendipity-oriented RSs may differ depending on the domain.

\noindent
\textbf{(3) Output of the LLM-as-a-Judge system. }
Across all the datasets and RSs, the Avg.\ Score remained in the 2.1--2.4 range.
This aligns with the ground truth $g$ distribution in the Serendipity-2018 dataset (Section~\ref{sec:prompt_selection_exp}), suggesting that the LLM-as-a-Judge system effectively reflects the serendipity assessments of real users.

\section{Conclusion} \label{sec:conclusion}
In this study, we propose a universal framework for serendipity evaluation using LLM-as-a-Judge.
Our experimental results suggest that the serendipity prediction accuracy can be improved by incorporating prompt designs that guide the inference process.
Moreover, evaluating several RSs with the proposed framework revealed that no single RS consistently delivers the best serendipitous performance, and that even classical RSs can achieve strong performance in certain cases.
In future studies, we plan to evaluate a broader range of datasets and RSs by using the proposed framework.

\begin{acks}
 This work was supported by JSPS KAKENHI Grant Numbers JP23K21724, JP23K24953, JP24K21410.
\end{acks}

\section*{GenAI Usage Disclosure}
In this study, we used two large language models (LLMs), GPT-4o-mini and Llama-3.1-70B Instruct, as research subjects in our experiments.
Details of their usage are described in Sections 4 and 5.
We did not use any GenAI tools for writing this manuscript.

\bibliographystyle{ACM-Reference-Format}
\bibliography{reference}

\end{document}